\documentclass[aps,prl,superscriptaddress,twocolumn,10pt]{revtex4-2}
\usepackage{amssymb}
\usepackage{cancel}
\usepackage{color,graphicx}
\usepackage{amsmath}
\usepackage{amsbsy}
\usepackage{amsthm}
\usepackage{bbm}
\usepackage{epsfig}
\usepackage{float}
\usepackage{graphicx}
\usepackage{subfigure}
\usepackage{dcolumn}
\usepackage{bbm}
\usepackage{color,epstopdf}
\usepackage{amscd}
\usepackage{amsfonts}  
\usepackage[]{amsmath}
\usepackage{amssymb}    
\usepackage{mathrsfs}
\usepackage{verbatim}
\usepackage[]{cases}
\usepackage{amsmath}
\usepackage{wasysym}
\usepackage[latin9]{inputenc}
\usepackage[T1]{fontenc}
\usepackage{mathtools}
\usepackage{todonotes}
\usepackage[normalem]{ulem}
\usepackage[colorlinks=true,linkcolor=blue,urlcolor=blue,citecolor=blue,pdfusetitle]{hyperref}
\usepackage{cleveref}
\usepackage{libertine}
\usepackage{hyperref}

\newcommand{\ket}[1]{\left| {#1} \right\rangle}

\definecolor{ao}{rgb}{0.0, 0.5, 0.0}
\definecolor{mypink}{rgb}{0.858, 0.188, 0.478}
\definecolor{mygreen}{rgb}{0.0, 0.5, 0.0}

\begin{document}

\title{Experimental signature of initial quantum coherence on entropy production}

\author{S. Hern\'andez-G\'omez}\thanks{These authors contributed equally to this work}
\affiliation{European Laboratory for Non-linear Spectroscopy (LENS), Universit\`a di Firenze, I-50019 Sesto Fiorentino, Italy}
\affiliation{Dipartimento di Fisica e Astronomia, Universit\`a di Firenze, I-50019, Sesto Fiorentino, Italy}
\affiliation{Istituto Nazionale di Ottica del Consiglio Nazionale delle Ricerche (CNR-INO), I-50019 Sesto Fiorentino, Italy}

\author{S. Gherardini}\thanks{These authors contributed equally to this work}
\affiliation{CNR-INO, Area Science Park, Basovizza, I-34149 Trieste, Italy}
\affiliation{European Laboratory for Non-linear Spectroscopy (LENS), Universit\`a di Firenze, I-50019 Sesto Fiorentino, Italy}

\author{A. Belenchia}\thanks{These authors contributed equally to this work}
\affiliation{Institut f\"ur Theoretische Physik, Eberhard-Karls-Universit\"at T\"ubingen, 72076 T\"ubingen, German}
\affiliation{School of Mathematics and Physics, Queen's University Belfast, Belfast BT7 1NN, United Kingdom}

\author{A. Trombettoni}
\affiliation{Department of Physics, University of Trieste, Strada Costiera 11, I-34151 Trieste, Italy}
\affiliation{SISSA and INFN, Sezione di Trieste, Via Bonomea 265, I-34136 Trieste, Italy}
\affiliation{CNR-IOM DEMOCRITOS Simulation Center and SISSA, Via Bonomea 265, I-34136 Trieste, Italy}

\author{M. Paternostro}
\affiliation{School of Mathematics and Physics, Queen's University Belfast, Belfast BT7 1NN, United Kingdom}

\author{N. Fabbri}
\affiliation{European Laboratory for Non-linear Spectroscopy (LENS), Universit\`a di Firenze, I-50019 Sesto Fiorentino, Italy}
\affiliation{Istituto Nazionale di Ottica del Consiglio Nazionale delle Ricerche (CNR-INO), I-50019 Sesto Fiorentino, Italy}

\begin{abstract}
We report the experimental quantification of the contribution to non-equilibrium entropy production that stems from the quantum coherence content in the initial state of a qubit exposed to both coherent driving and dissipation. Our experimental demonstration builds on the exquisite experimental control of the spin state of a nitrogen-vacancy defect in diamond and is underpinned, theoretically, by the formulation of a generalized fluctuation theorem designed to track the effects of quantum coherence. Our results provide significant evidence of the possibility to pinpoint the genuinely quantum mechanical contributions to the thermodynamics of non-equilibrium quantum processes. 
\end{abstract}

\maketitle

The irreversible character of most physical processes is, apparently, at odds with the inherent reversibility of the fundamental laws of physics. The way time-reversible quantum laws governing the interactions of microscopic systems gives rise to the irreversible nature of macroscopic phenomena is a very open field of investigation~\cite{LandiRMP2021}. In this regard, a breakthrough has been provided by the extension of the second law of thermodynamics into the quantum realm through the so-called fluctuation theorems~\cite{JarzynskiPRL1997,PhysRevE.60.2721,CampisiRMP2011,RevModPhys.81.1665}. A celebrated instance of this is the integral fluctuation theorem, which stems from Jarzynski's identity~\cite{JarzynskiPRL1997} and Crooks' relation~\cite{PhysRevE.60.2721}, and connects the non-equilibrium energy fluctuation statistics of unital processes with the corresponding free-energy changes. Operationally, the standard approach to the quantification of energy and entropy fluctuations in non-equilibrium contexts is the use of the celebrated two-point measurement (TPM) scheme~\cite{TalknerPRE2007}, which requires two projective measurements, at the beginning and at the end of the dynamical process under scrutiny~\cite{GherardiniQSTreconstructing,PhysRevX.8.031037}. 

Despite the clear success of the TPM scheme, evidenced by  successful experimental verification in nuclear magnetic resonance~\cite{Batalhao2014,Batalhao2015}, trapped-ion~\cite{An15,Smith18,Xiong18}, superconducting-qubit~\cite{Zhang18}, nitrogen-vacancy (NV) centers~\cite{HernandezPRR2019,HernandezNJP2021} and linear optics settings~\cite{cimini2020experimental,Ribeiro20,Aguilar21x}, the scheme has significant limitations when considering the role played by quantum features in the statistics of energy fluctuations. In fact, any quantum coherence in the initial state of the system, and expressed in the measurement basis, is washed away. As a result, the system dynamics following the first measurement is strongly affected. This has motivated recent efforts aimed at modifying the TPM scheme to take into account the presence of quantum features, particularly coherence, in non-equilibrium processes~\cite{AllahverdyanPRE2014,DeffnerPRE2016,LostaglioPRL2018,SantosnpjQI2019,PhysRevX.9.031029,MicadeiPRL2020,PhysRevLett.127.180603,Sone2020,LostaglioKirkwood2022,HernandezQPsArXiv2022}. Ref.~\cite{Ste_Ale_arXiv2021} introduced an end-point measurement (EPM) approach, where the initial statistics of energy fluctuations is inferred from the knowledge of the initial state and the Hamiltonian of the system. 

Here, we show the intrinsic operational nature of the EPM approach by considering both the detailed and the integral form of the corresponding fluctuation theorem, and using them to characterise experimentally the entropy production associated to quantum coherence in an open quantum system's states. In particular, we experimentally make use of a qubit encoded in the spin of an NV center in diamond, subjected to both a pulsed driving and environmental effects~\cite{HernandezPRR2019,HernandezNJP2021,HernandezPRXQ2022}. We thus observe a significant increase of the irreversibility of the resulting dissipative map that only originates from the presence of quantum coherence in the initial state of the NV center. Moreover, we show that measuring such a quantity provides a tight bound for the average heat exchanged by the system with the environment. 

While being valid in principle for arbitrary dynamics, our results establish NV centers as valuable platforms for the exploration of energetics at the quantum level, thus enlarging the already prominent domain of their applications in quantum technologies~\cite{Maze2008,Rondin14,Degen17,HernandezFiP2021}.

\textbf{EPM-based fluctuation theorem.}--
We start by briefly reviewing the EPM scheme as introduced in Ref.~\cite{Ste_Ale_arXiv2021}. We thus consider a quantum system subjected to a completely-positive trace preserving (CPTP) map $\Phi$. Let $\rho_0 = \mathcal{P}+\chi$ be its initial state, which we have decomposed in its diagonal part $\mathcal{P}$ (expressed in the basis of the initial Hamiltonian $H_{t_i}$) and the traceless component $\chi$ that accounts for the quantum coherence. The EPM scheme prescribes to perform a single energy measurement, at the end of the process, and to associate to it the stochastic variables $\Delta E_{\rm i,f} \equiv E^{{\rm fin}}_{\rm f} - E^{{\rm in}}_{\rm i}$ that encodes the energy fluctuations during the open dynamics. Here, $E_{\rm k}^{{\rm in}({\rm fin})}$ denotes the eigenvalues of the initial (final) Hamiltonian. We also introduce the probability distribution 
\begin{equation}\label{eq:pdf}
    p_{\rm EPM}(\Delta E_{\rm i,f}) = 
    p(E^{{\rm in}}_{\rm i})p(E^{{\rm fin}}_{\rm f})={\rm tr}(\Pi^{\rm in}_{\rm i}\rho_0)
    {\rm tr}(\Pi^{\rm fin}_{\rm f}\Phi(\rho_0)),
\end{equation}
where $\Pi^{{\rm in}({\rm fin})}_{\rm j}$ is the projector on the $j$-th  initial (final) energy eigenstate. Computing the characteristic function of the probability distribution in Eq.~\eqref{eq:pdf} leads immediately to an integral fluctuation theorem. In fact, let us take ${\cal P}=\rho_{\rm{th}}\equiv Z_{i}^{-1}\exp[-\beta H_{t_i}]$, i.e., a thermal state of the initial Hamiltonian with inverse temperature $\beta$ and partition function $Z_{i} \equiv {\rm tr}(\exp[-\beta H_{t_i}])$. One can then show that 
\begin{equation}\label{eq:GFT}
    \langle e^{-\beta(\Delta E - \Delta F)}\rangle = d\left[{\rm tr}\left(\rho^{\rm fin}_{\rm th}\,\Phi(\rho^{\rm in}_{\rm th})\right)+{\rm tr}\left(\rho^{\rm fin}_{\rm th}\,\Phi(\chi)\right)\right],
\end{equation}
where $\rho^{\rm fin/in}_{\rm th} \equiv e^{-\beta H_{t_{f/i}}}/Z_{f/i}$ with $Z_{f} \equiv {\rm tr}(\exp[-\beta H_{t_f}])$, and $d$ denoting the dimension of the system's Hilbert space. The second term in the right-hand-side of Eq.\,(\ref{eq:GFT}) showcases the contribution coming from the initial quantum coherence, while the first term represents a classical deviation from the Jarzynski's equality. Such deviation would be present also in the absence of initial coherence due to the non-linearity of the EPM's probability distribution for convex combination of states~\cite{Ste_Ale_arXiv2021}. 

Consider now a map $\Phi$ that admits a (non-singular) fixed point $\rho^*=\Phi(\rho^*)$. Using the results in~\cite{PhysRevA.77.034101}, we can define the corresponding backwards dynamics [see~\cite{SM} (Part~A)]. Then, one can compare the joint probabilities ${\rm P}_{\Gamma}({\rm i,f}) = p(E_{\rm i}^{\rm in})p(E_{\rm f}^{\rm fin}) \equiv p_{\rm i}^{\rm in}p_{\rm f}^{\rm fin}$ and ${\rm P}_{\widetilde{\Gamma}}({\rm f,i}) = {\rm tr}(\Pi^{\rm fin}_{\rm f}\rho^{\rm in}_{\rm B}){\rm tr}(\Pi^{\rm in}_{\rm i}\widetilde{{\Phi}}(\rho^{\rm in}_{\rm B})) \equiv \widetilde{p}^{\rm in}_{\rm f}\widetilde{p}^{\rm fin}_{\rm i}$ for measuring the energy of the system in the forward and backward trajectories, $\Gamma$ and $\widetilde{\Gamma}$ respectively. Note that, in ${\rm P}_{\widetilde{\Gamma}}({\rm f,i})$, $\rho_{\rm in}^{\rm B}$ denotes the initial state of $\widetilde{\Gamma}$, and $\widetilde{\Phi}$ is the time-reversed map~\cite{SM} (Part~A).

By extending the derivation of the Jarzynski equality~\cite{JarzynskiPRL1997} by means of the Crooks' formalism~\cite{PhysRevE.60.2721}, we consider the case in which \emph{(i)} the initial quantum state of the forward dynamics is $\rho_{0}=\rho_{\rm th}^{\rm in}(\beta)+\chi$, i.e., $\rho_{0}$ is written as the sum of a thermal state at inverse temperature $\beta$ and the traceless component $\chi$ encoding the initial coherence in the energy basis, and \emph{(ii)} the initial state of the backward quantum dynamics is the state that is thermal in the final Hamiltonian at the same inverse temperature $\beta$, $\rho^{\rm in}_{\rm B}=Z_{f}^{-1}\exp[-\beta H_{t_f}]$.
These assumptions allow to write the balance equation 
\begin{equation}\label{eq:mimick_Jarzynski_3_over_2}
    \frac{{\rm P}_{\Gamma}(\rm i,f)}{{\rm P}_{\widetilde{\Gamma}}(\rm f,i)} = \exp\left[\beta(\Delta E_{\rm i,f}-\Delta F)+\Delta\sigma_{\rm i,f}+\Delta\Sigma_{\rm i,f}\right],
\end{equation}
where
\begin{align}
\label{Sigma}
    &\Delta\sigma_{\rm i,f} \equiv \ln\frac{p_{\rm f}^{\rm fin}(\rho^{\rm in}_{\rm th}(\beta))}{\widetilde{p}^{\rm fin}_{\rm i}(\widetilde{\Phi}(\rho^{\rm in}_{\rm B}))},~\Delta\Sigma_{\rm i,f} \equiv \ln\left[1+\frac{p^{\rm fin}_{\rm f}(\chi)}{p^{\rm fin}_{\rm f}(\rho_{\rm th}^{\rm in}(\beta))}\right]
\end{align}
and $p^{\rm fin}_{\rm f}(\chi) \equiv {\rm tr}(\Pi^{\rm fin}_{\rm f}\Phi(\chi))$ (cf.~Part~A of Ref.~\cite{SM}). In this way, averaging Eq.~(\ref{eq:mimick_Jarzynski_3_over_2}) over the forward probability distribution, one obtains 
\begin{equation}\label{eq_integral_fluct_relation}
    \left\langle e^{-\beta\Delta E-(\Delta\sigma + \Delta\Sigma)}\right\rangle_{\Gamma} = e^{-\beta\Delta F}\,.
\end{equation}
Eq.~\eqref{eq:mimick_Jarzynski_3_over_2} and Eq.~\eqref{eq_integral_fluct_relation} are the \emph{detailed} and the \emph{integral} forms of the EPM's fluctuation theorem, respectively.

\begin{figure}[t!]
\centering
{\bf (a)}\hskip5cm{\bf (b)}\\
\includegraphics[width=1\columnwidth]{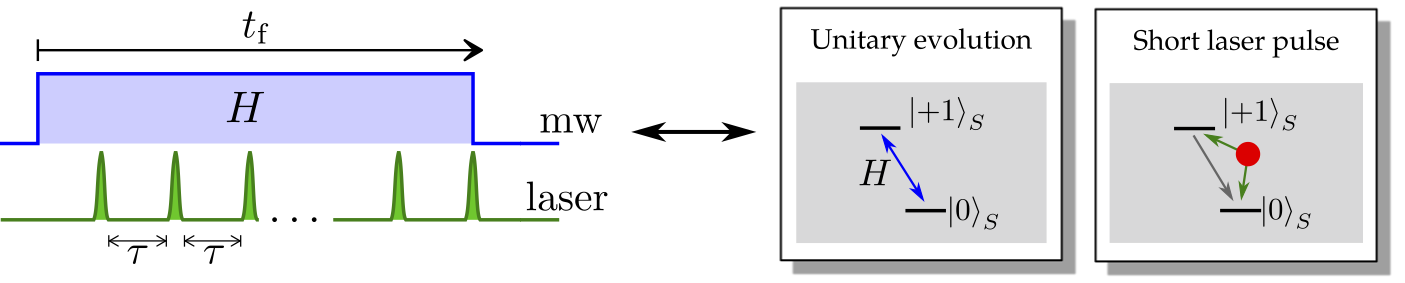}
\caption{
Scheme of the protocol applied to the NV spin qubit. The dynamics stems from a combination of coherent driving of the qubit (near-resonance microwave (mw) radiation) and a train of short laser pulses that ``open'' the system [panel {\bf (a)}]. Specifically, the coherent drive couples the states $\ket{0}_S$ and $\ket{+1}_S$. Instead, the short laser pulses act as a dissipation, by first projecting the system into the $S_z$ eigenstates and then optically pumping part of the populations towards the $\ket{0}_S$ state [panel {\bf (b)}]. 
}
\label{fig:applied_protocol}
\end{figure}

It is crucial to notice that, thanks to the assumption \emph{(ii)}, the quantity $\Delta\Sigma_{\rm i,f}$ depends only on the forward dynamics and it satisfies a fluctuation theorem by its own (cf.~Part~B of Ref.~\cite{SM}), i.e., 
\begin{equation}\label{SigmaFT}
    \left\langle e^{-\Delta\Sigma}\right\rangle_{\Gamma}=1 \,.
\end{equation}
Thus, resorting to the Jensen's inequality, we have $\left\langle\Delta\Sigma\right\rangle\geq 0$. $\Delta\Sigma$ encodes the entropic contribution of the initial quantum coherence of the system, and we thus identify it as the \textit{coherence-affected irreversible entropy production} for a non-equilibrium dynamical process. At the same time, the quantity $\Delta\sigma$ represents a completely classical contribution to the entropy production that comes from adopting the EPM formalism, namely from the extra uncertainty implied by the factorization condition in Eq.~\eqref{eq:pdf} (see also discussion in~\cite{Ste_Ale_arXiv2021}).

\textbf{The experimental system.}--
We consider the spin qubit associated to a negatively charged NV center --a localized impurity in a diamond lattice based on a nitrogen substitutional atom next to a vacancy-- which forms an electronic spin $S = 1$ in its orbital ground state~\cite{Gruber97,Jelezko04,Doherty13,Aharonovich14}. 
A magnetic bias field aligned with $S_z$ removes the degeneracy of the spin eigenstates, so as to allow for the selective coherent manipulation of the transition $|0\rangle_S \leftrightarrow \ket{+1}_S$. The Hamiltonian $H$ of this effective two-level system is determined by a continuous nearly resonant microwave field and, in the frame rotating at the microwave frequency, the Hamiltonian is $H = \hbar\omega(\cos\alpha\,\tilde{\sigma}_{z} - \sin\alpha\,\tilde{\sigma}_x)/2$, where $\omega=\sqrt{\Omega^2+\delta^2}$ and $\tan\alpha = -\Omega/\delta$, $\Omega$ denotes the bare Rabi frequency and $\delta\in [0, \Omega]$ is the microwave detuning. Note that we have used the tilde for the Pauli matrices in view of the change of basis to the Hamiltonian eigenstates, i.e., $\{|0\rangle,|1\rangle\} \equiv \{\cos\frac{\alpha}{2}\,\ket{0}_{S} - \sin\frac{\alpha}{2}\,\ket{+1}_S, \sin\frac{\alpha}{2}\,\ket{0}_S + \cos\frac{\alpha}{2}\,\ket{+1}_S\}$ with eigenvalues $\pm \hbar\omega/2$. In this new basis, the Hamiltonian becomes $H=\omega\sigma_{z}/2$, where $\hbar$ is set to $1$ from here on.

The qubit is governed by an alternated sequence of unitary and non-unitary (controlled-dissipative) dynamics, as follows.
The system is repeatedly subjected to a sequence of pulses, occurring regularly at time intervals $\tau$. Among two consecutive pulses, the evolution of the NV center is unitary and described by the operator $U \equiv \exp[-iH\tau]$. 
As depicted in Fig.~\ref{fig:applied_protocol}, the NV spin is subjected to open dynamics due to its interaction with a train of short laser pulses with a duration $t_L$ that is much shorter than the characteristic time-scale of the unitary dynamics ($t_L\ll2\pi/\omega $). The short laser pulses trigger cycles of spin preserving and non-preserving transitions between different orbital levels~\cite{HernandezPRR2019}. This entails non-unitary dynamics that project the state of the system into the eigenstates of $\tilde{\sigma}_z$ and partially transfer the spin population to $|0\rangle_S$. Such \emph{spin amplitude damping} along the $\tilde{\sigma}_z$ axis, also known as optical pumping, is caused by the spin non-preserving transitions, and can be modelled as a controlled dissipative channel toward $|0\rangle_S$~\cite{HernandezPRR2019,HernandezPRXQ2022}. The overall dynamics takes the NV center into an asymptotic fixed point $\rho^*$ that we can use to define the backward dynamics~\cite{SM} (Parts~A,C).

In the experiments shown throughout the letter, we set $\alpha = \pi/4$ (i.e., $\delta=-\Omega$), $\tau\omega\simeq (2\pi)0.9$, and $\tau=190$~ns.

\textbf{Coherence-affected entropy production.}--
We now show the results obtained with the NV qubit subjected to the dissipative dynamics introduced above. Here, our aim is to characterize the thermodynamic role of initial quantum coherence in terms of the coherence-affected entropy production. 
\begin{figure}[b!]
\centering
\includegraphics[width=\columnwidth]{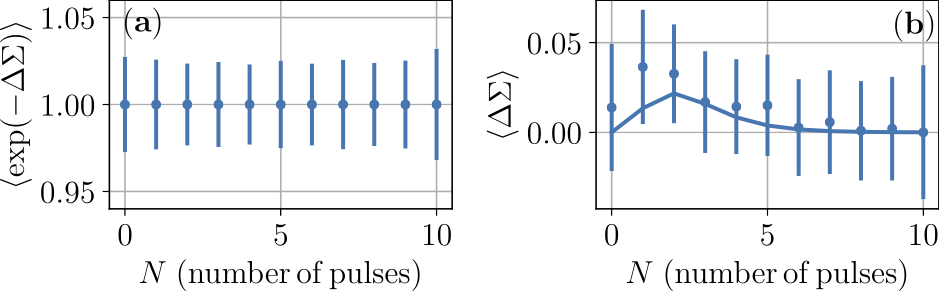}
\caption{Panel {\bf (a)}: Experimental verification of the fluctuation theorem $\left\langle e^{-\Delta\Sigma}\right\rangle_{\Gamma}=1$, Eq.~\eqref{SigmaFT}, for the coherence-affected irreversible entropy production, as a function of the number $N$ of pulses used to drive the dynamics of the NV center. Panel {\bf (b)}: Average experimental coherence-affected entropy production as a function of $N$ (blue circles). In both panels, the experimental data are plotted against the predictions from the numerical simulations that are obtained by taking $|+\rangle_{y}=(\ket{0}+i\ket{1})/\sqrt{2}$ (i.e., $\rho_0$ with $p=0$) as the initial quantum state. For such an initial state, the coherence-affected entropy production is nearly extremal.}
\label{SigmaandFT}
\end{figure}
We have thus performed a series of experiments to determine both the EPM probability distribution for the energy fluctuations statistics as well as the usual TPM one.

At the beginning of each experimental realization, the electronic spin is initialized into the $|0\rangle_S$ eigenstate of the spin operator $S_z$ via optical spin pumping under long laser excitation. Then, the system is brought in each of the four different pure states that correspond to the eigenvectors of $\sigma_z$ and $\sigma_y$, by applying rotation gates (on-resonant microwave pulse) to the state $|0\rangle_S$. After $n$ short laser pulses, with $n\in [0,N]$, we measure the energy of the system in the Hamiltonian basis $\sigma_z$. To achieve this, we first apply another rotation gate (on-resonant microwave pulse) such that $\sigma_z \rightarrow \tilde{\sigma}_z$, and then we measure the photo-luminescence (PL). The PL intensity determines the probability for the system to be in the eigenstates $|0\rangle_S$ or $|+1\rangle_S$.
The experiment is repeated $10^6$-times from the beginning for each value of the pulses number. In this way, all the probabilities required to obtain the EPM and TPM statistics are obtained.

As we collect data from several experiments (in which the system is initialized in one of the four pure qubit-states $\{|0\rangle,|1\rangle,|+\rangle_y,|-\rangle_y\}$), it is a matter of data processing to compute the EPM and TPM statistics for every classical mixture of such states~\cite{SM} (Part~C). In particular, looking at a Bloch sphere representation for the qubit, this implies that we are able to obtain the energy statistics of our quantum process initialized in one of the states, which are included in the $y-z$ equatorial plane of the Bloch sphere and provided by a convex combination of the states $|0\rangle,|1\rangle,|+\rangle_y,|-\rangle_y$. Specifically, in the following we show experimental results corresponding to the initial states (expressed in the Hamiltonian basis)
\begin{equation}\label{eq:initial_state_exp}
\rho_0 = \frac{1}{2}\begin{pmatrix} 1+p & -(1-p) i \\ (1-p) i & 1-p
\end{pmatrix}
\qquad (p\in [0,1]) 
\end{equation} 
such that $\rho_0$ is the convex mixture of $|1\rangle\!\langle 1|$ and $|+\rangle_y\!\langle +|$ with probability $p$ and $1-p$, respectively.

The first quantity we are interested in characterising experimentally is the coherence-affected entropy production encoded in the average of $\Delta\Sigma$ as given in Eq.~\eqref{Sigma}. Note that this average is defined solely in terms of the forward trajectory probability, so that we can fully characterize it by resorting to the experimental data acquired during the forward dynamics. The results are shown in Fig.~\ref{SigmaandFT}: in the left panel we show the experimental verification of the fluctuation theorem in Eq.~\eqref{SigmaFT}. Instead, in the right panel, the behaviour of $\langle\Delta\Sigma\rangle$ is plotted as a function of the number of laser pulses. While the error bars are quite large, it can be observed how the corresponding experimental points nicely follow the theoretical predictions and how the experimental data show a positive coherence-affected entropy production. 

\begin{figure}[b!]
\centering
\includegraphics[width=\columnwidth]{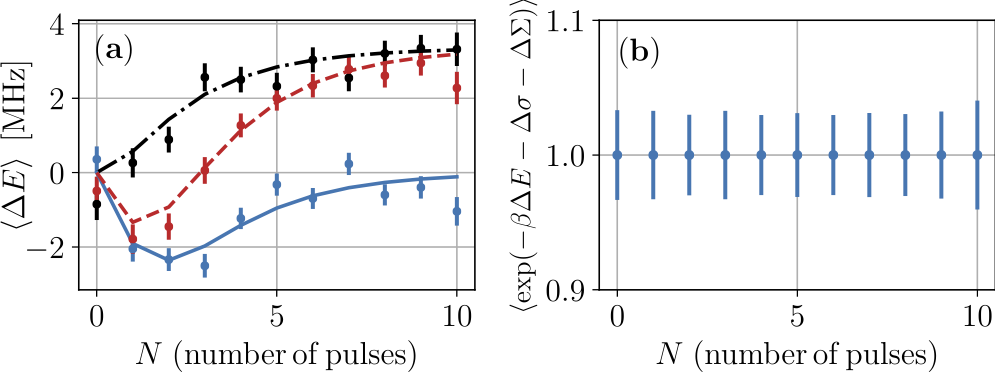}
\caption{Panel {\bf (a)} shows the three quantities in Eq.~\eqref{energies}. The dashed red curve stands for $\langle\Delta E\rangle_{\rm EPM}$, the dot-dashed black one for $\langle\Delta E\rangle_{\rm TPM}$, and the solid blue line shows the contribution of quantum coherence to the exchanged heat $\sum_{\rm f}{\rm tr}[\Pi^{\rm fin}_{\rm f}\chi] E_{\rm f}$. The curves are obtained for an initial state parameterized as in Eq.~\eqref{eq:initial_state_exp} with $p=0.2$. This choice guarantees a good visibility of all the contributions. Panel {\bf (b)}: Experimental verification of the integral form of the EPM fluctuation theorem in Eq.~\eqref{eq_integral_fluct_relation}. In this panel, we have taken Eq.~\eqref{eq:initial_state_exp} with $p=0.38$.}
\label{fig:Average_energies}
\end{figure}

Another quantity that can be investigated directly from the available data on the forward dynamics is the average of the energy fluctuations $\Delta E$. In this regard, it is worth noting that this quantity is identified with the average work when considering time-dependent unitary processes like in the Jarzynski's original work \cite{JarzynskiPRL1997}. In our case, as the Hamiltonian is time-independent, we can unambiguously interpret this quantity as the average \emph{heat} that the system exchanges with its environment due to the open dynamics to which the NV center is subjected. The average of the stochastic variable $\Delta E_{\rm i,f}=E_{\rm f}-E_{\rm i}$ in the EPM approach is related to the TPM scheme via the following relation:
\begin{equation}\label{energies}
    \langle\Delta E\rangle_{\rm EPM} = \langle \Delta E\rangle_{\rm TPM} + \sum_{\rm f}{\rm tr}(\Pi^{\rm fin}_{\rm f}\chi) E_{\rm f}\,.
\end{equation}
The second term on the right-hand side of Eq.~(\ref{energies}) represents a contribution to the average energy-change ascribable to the quantum coherence of the initial state, which is not deleted by a first energy measurement of $\rho_0$. Fig.~\ref{fig:Average_energies} displays the three quantities entering in Eq.~\eqref{energies}, by comparing theoretical expectations with the experimental results. It is shown that the experimental data are able to discern the coherence contribution to the heat exchanged between the system and the environment due to the pulsed dynamics. Among the protocols that allows to account for quantum features in energy fluctuations~\cite{AllahverdyanPRE2014,DeffnerPRE2016,LostaglioPRL2018,SantosnpjQI2019,MicadeiPRL2020,PhysRevLett.127.180603,Sone2020,Ste_Ale_arXiv2021,LostaglioKirkwood2022}, the EPM scheme requires only a final energy measurement at the time $t$ and does not require any knowledge of the quantum map $\Phi$ that models the forward dynamics. This makes the method particularly suitable for application in open quantum systems.

Finally, we have also verified the validity of the full EPM fluctuation theorem in Eq.~\eqref{eq_integral_fluct_relation} as well as its consequences for the expectation values of the involved thermodynamic quantities. In principle, this requires having access to the backward trajectories of the system, so as to characterize $\Delta\sigma_{\rm i,f}$ in Eq.~\eqref{Sigma}. This can be problematic due to the presence of non-unitary dynamics: while the backward trajectories can be easily simulated numerically, implementing them at the experimental level is not currently possible with our set-up. However, for the range of experimental parameters and the choice of the initial state of the backward process, the backward dynamics are such that $\widetilde{p}^{\rm fin}_{\rm j}(\widetilde{\Phi}(\rho^{\rm in}_{\rm B}))=p_{\rm j}^{\rm fin}(\rho^{\rm in}_{\rm th}(\beta))$. Therefore, by assuming that this property holds and thus that we can estimate $\langle\Delta\sigma\rangle$ from the sole data of the forward trajectories, in Fig.~\ref{fig:Average_energies} {\bf (b)} we show the experimental verification of the integral form of the EPM fluctuation theorem in Eq.~\eqref{eq_integral_fluct_relation}. 
\begin{figure}[t!]
\centering
\includegraphics[width=\columnwidth]{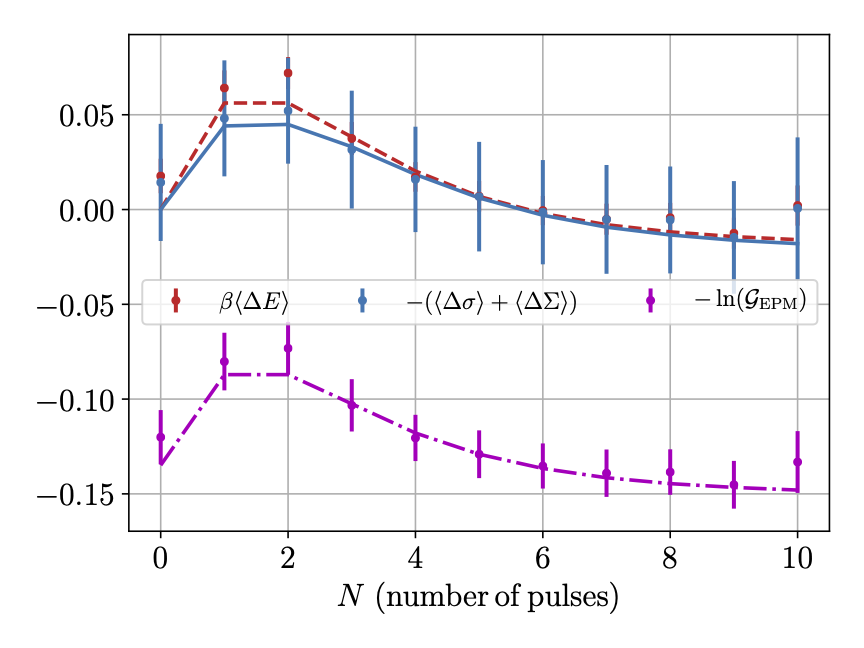}
\caption{
The dashed red curve corresponds to the average EPM energy-change $\beta\langle\Delta E\rangle$. The dot-dashed magenta and solid blue solid curves show the behavior of the right-hand-sides of the inequalities in  Eq.~\eqref{JFT}. The initial state is taken as in Eq.~\eqref{eq:initial_state_exp} with $p=0.38$.
}
\label{fig:energetic_bound2}
\end{figure}

The application of the Jensen inequality to Eqs.~\eqref{eq:GFT} and~\eqref{eq_integral_fluct_relation} leads to 
\begin{equation}
    \begin{aligned}
    \beta\langle\Delta E\rangle\geq
    \begin{cases}
    -\ln({\cal G}_{\rm EPM}),\\
    - (\langle\Delta\sigma\rangle + \langle\Delta\Sigma\rangle)
    \end{cases}
    \label{JFT}
\end{aligned}
\end{equation}
with ${\cal G}_{\rm EPM}=2{\rm tr}\left(\rho^{\rm fin}_{\rm th}\Phi(\rho^{\rm in}_{\rm th}+\chi)\right)$ the EPM characteristic function~\cite{Ste_Ale_arXiv2021}. In Fig.~\ref{fig:energetic_bound2}, the theoretical expectations of these quantities are compared with the corresponding experimental results. For the initial state $\rho_0$ considered for this figure (i.e., the state in Eq.~\eqref{eq:initial_state_exp} with $p=0.38$), one can observe that the results obtained by using the characteristic function of the EPM approach are quite distinct from the ones coming from the integral fluctuation theorem \eqref{eq_integral_fluct_relation}. Specifically, for the number of pulses explored experimentally and for the initial states accessible from the experimental data, the bound on the average energy-change derived from the integral fluctuation theorem is \emph{tighter} than the one resulting from the EPM characteristic function. In addition, we can also see in Fig.~\ref{fig:energetic_bound2} that the inequality in Eq.~\eqref{JFT} is almost saturated, so that the EPM average energy offers even a good estimate of the sum $\langle\Delta\Sigma+\Delta\sigma\rangle$ in the regime experimentally analyzed. 

\textbf{Conclusions.}--
We have used the EPM approach, for the characterization of energy fluctuations arising from a non-equilibrium process, with the aim to quantify the contribution to the entropy production that is originated by the presence of quantum coherence in the initial density matrix of the quantum system under scrutiny. The operational nature of our approach has enabled a successful experimental quantification of such coherence-affected entropy production in a solid state platform where an NV spin qubit undergoes controlled dissipative dynamics. 

Our study grounds the EPM approach as a powerful framework for the assessment of quantum coherence in the energetics of quantum systems and devices. Specifically, our findings add a crucial ingredient for the analysis of the thermodynamic role of quantum coherence and allow the characterization of the newly-introduced entropy production due to quantum coherence.

\textbf{Acknowledgments.}--
We acknowledge support from the European Union's Horizon 2020 FET-Open project TEQ (766900), the Leverhulme Trust Research Project Grant UltraQuTe (grant RGP-2018-266), the Royal Society Wolfson Fellowship (RSWF/R3/183013), the UK EPSRC (grant EP/T028424/1), the Department for the Economy Northern Ireland under the US-Ireland R\&D Partnership Programme, the Deutsche Forschungsgemeinschaft (DFG, German Research Foundation) project number BR 5221/4-1, the Blanceflor Foundation through the project ``The theRmodynamics behInd thE meaSuremenT postulate of quantum mEchanics (TRIESTE)'', the CNR-FOE-LENS-2020, and from the European Commission under GA n.\,101070546--MUQUABIS.

\bibliography{biblio}

\clearpage
\onecolumngrid
\begin{center}
{\bf Supplementary Material:\\
Experimental signature of initial quantum coherence on entropy production}
\end{center}

\setcounter{figure}{0}
\setcounter{table}{0}
\setcounter{page}{1}
\makeatletter
\renewcommand{\theequation}{S\arabic{equation}}
\renewcommand{\thefigure}{S\arabic{figure}}
\renewcommand{\bibnumfmt}[1]{[S#1]}
\renewcommand{\citenumfont}[1]{S#1}

\appendix

\section{Part A: Derivation of the detailed fluctuation theorem}\label{app:FTs}

\renewcommand{\theequation}{A-\arabic{equation}}
\setcounter{equation}{0}

Let us consider a CPTP map, written as $\Phi(\bullet)=\sum_{\alpha}K_{\alpha}\bullet K_\alpha^\dag$ with $\{K_\alpha\}$ the set of Kraus operators, that allows a (non-singular) fixed point $\rho^*$ such that $\rho^*=\Phi(\rho^*)$. Following Crooks' formalism~\cite{PhysRevE.60.2721}, we define the time-reversed map  
\begin{equation}
\widetilde{\Phi}(\bullet)=\sum_{\alpha}[(\rho^{*})^{1/2} K^\dag_{\alpha}(\rho^*)^{-1/2}]\bullet[(\rho^{*})^{-1/2} K^\dag_{\alpha}(\rho^*)^{1/2}],
\end{equation}
where $\widetilde{K}_\alpha \equiv (\rho^{*})^{1/2} K^\dag_{\alpha}(\rho^*)^{-1/2}$.

Afterwards, in accordance with the EPM formalism~\cite{Ste_Ale_arXiv2021}, we can introduce the expression of the joint probabilities ${\rm P}_{\Gamma}(E^{\rm in}_{\rm n},E^{\rm fin}_{\rm m}) \equiv {\rm P}_{\Gamma}(n,m)$ and ${\rm P}_{\widetilde{\Gamma}}(E^{\rm fin}_{\ell},E^{\rm in}_{k}) \equiv {\rm P}_{\widetilde{\Gamma}}(\ell,k)$, respectively, for the quantum trajectories $\Gamma$ and $\widetilde{\Gamma}$. Let us observe that the latter is obtained by reversing the arrow of time by means of a transformation that obeys the time-reversal symmetry. One has
\begin{equation}
{\rm P}_{\Gamma}(n,m) = {\rm tr}\left(\Pi_{n}^{\rm in}\rho_{0}\right){\rm tr}\left(\Pi_{m}^{\rm fin}\rho^{\rm fin}\right) \equiv p_{n}^{\rm in}p_{m}^{\rm fin} 
\end{equation}
and
\begin{equation}
{\rm P}_{\widetilde{\Gamma}}(\ell,k) = {\rm tr}\left(\Pi_{\ell}^{\rm fin}\rho_{\rm B}^{\rm in}\right){\rm tr}\left(\Pi_{k}^{\rm in}\rho_{\rm B}^{\rm fin}\right) \equiv \widetilde{p}_{\ell}^{\rm in}\widetilde{p}_{k}^{\rm fin} 
\end{equation}
where, we recall, $\rho^{\rm fin} \equiv \Phi[\rho_0]$ and $\rho_{\rm B}^{\rm fin} \equiv \widetilde{\Phi}[\rho_{\rm B}^{\rm in}]$ with $\rho_{\rm B}^{\rm in}$ denoting the initial state of the backward process and, without loss of generality, we have assumed that each projector on the energy eigenstates is invariant to the application of the time-reversal transformation. As a result, the expression of the \emph{detailed balance equation} for the trajectories $\Gamma$ and $\widetilde{\Gamma}$ is formally equal to
\begin{equation}\label{detailed_be_app}
    \frac{{\rm P}_{\Gamma}(n,m)}{{\rm P}_{\widetilde{\Gamma}}(\ell,k)} = \frac{p^{\rm in}_{n}p^{\rm fin}_{m}}{\widetilde{p}^{\rm in}_{\ell}\widetilde{p}^{\rm fin}_{k}}\,.
\end{equation}
Notice that, for a generic open system dynamics, the state originating from applying a time-reversal transformation on $\rho^{\rm fin}$ is at most in the neighborhood of the initial state $\rho_{\rm B}^{\rm in}$ of the backward process. Such a discrepancy along the time-reversal trajectory $\widetilde{\Gamma}$ in the space of density operators necessarily entails quantum entropy production. The latter is the main signature of thermodynamic irreversibility due to both the open system dynamics and the procedure we use to characterize energy-change fluctuations. 

Before proceeding, it is worth observing that if \textit{(i)} we apply the EPM scheme for characterising energy fluctuations and \textit{(ii)} we choose the initial state of the backward process as the time-reversal of the final state of the forward process, then $\widetilde{p}^{\rm in}_{\ell} = p^{\rm fin}_{m}$ for $\ell=m$. As a matter of fact,
\begin{equation}\label{eq:equality_probs}
    \widetilde{p}^{\rm in}_{\ell} = {\rm tr}\left(\rho^{\rm in}_{\rm B}\Pi^{\rm fin}_{\ell}\right) = {\rm tr}\left(\Theta\rho^{\rm fin}\Theta^{\dagger}\Theta\Pi^{\rm fin}_{\ell}\Theta^{\dagger}\right) = {\rm tr}\left(\rho^{\rm fin}\Pi^{\rm fin}_{\ell}\right) = p^{\rm fin}_{\ell}
\end{equation}
where $\widetilde{(\bullet)} \equiv \Theta(\bullet)\Theta^{\dagger}$ with $\Theta$ denoting the time-reversal operator, which by construction is anti-unitary, i.e., it is an anti-linear operator and satisfies the relations $\Theta^{\dagger}\Theta = \Theta\Theta^{\dagger} = \mathbbm{I}$. In this context, we obtain the fluctuation relation
\begin{equation}\label{special0}
    \frac{{\rm P}_{\Gamma}}{{\rm P}_{\widetilde{\Gamma}}}(n,m) =e^{-\Delta\overline{\sigma}^{\rm in}_{n}}
\end{equation} 
where we have identified 
\begin{equation}
\Delta\overline{\sigma}^{\rm in}_{n} = 
-\ln\frac{ {\rm tr}\left(\rho_0\Pi^{\rm in}_{n}\right) }{ {\rm tr}\left(\widetilde{\Phi}(\rho^{\rm in}_{\rm B})\Pi^{\rm in}_{n}\right) }\,.
\end{equation}
This relation encodes information only on the initial stochastic quantum entropy production due to the open system dynamics. In fact, for the special case in which the dynamics is \emph{unitary}, and under the assumption of micro-reversibility, i.e., $\Theta\,\mathcal{U}(\lambda)\Theta^\dag = \mathcal{U}^{\dagger}(\widetilde{\lambda})$ with $\lambda(t)$ generic time-dependent transformation such that the system Hamiltonian is invariant under time reversal, one finds that $e^{-\Delta\overline{\sigma}^{\rm in}_{n}}=1\,\,\forall n$, i.e., 
\begin{equation}\label{eq_unitary_case}
    {\rm P}_{\Gamma}(n,m) = {\rm P}_{\widetilde{\Gamma}}(n,m) \,.
\end{equation}
Note also that, in determining the integral form of Eq.~\eqref{special0}, one gets $\left\langle e^{\Delta\overline{\sigma}^{\rm in}}\right\rangle_{\Gamma} =1$ with 
\begin{equation}
    \left\langle\Delta\overline{\sigma}^{\rm in}\right\rangle_{\Gamma} = \sum_{n}p^{\rm in}_{n}\ln\frac{\widetilde{p}^{\rm fin}_{n}}{p^{\rm in}_{n}} = -S(p||\widetilde{p})
\end{equation}
where, we recall, $p^{\rm in}_{n}$ and $\widetilde{p}^{\rm fin}_{n}$ are the probabilities to measure the $n$-th energy value of the system, respectively, at the initial and final time instants of the forward and backward process. Here, $S(q||p)$ denotes the \emph{classical relative entropy} between the two probability distributions $q$ and $p$, and thus naturally corresponds to a measure of how far is the final state of the inverse quantum dynamics from the initial quantum state.

Coming back to the derivation of Eq.~\eqref{eq:mimick_Jarzynski_3_over_2} of the main text, let us assume \emph{(i)} that the initial quantum state $\rho_{0}$ of the forward dynamics has thermal populations but also non-zero coherence terms (in the system energy basis), and \emph{(ii)} that the initial quantum state $\rho^{\rm in}_{\rm B}$ of the backward quantum dynamics has thermal populations (with respect to the final Hamiltonian) and once again non-vanishing off-diagonal elements. In other terms,
\begin{align}
    & \rho_{0} = \rho_{\rm th}^{\rm in}(\beta) + \chi^{\rm in} \equiv \frac{e^{-\beta H_{t_i}}}{Z_{i,\beta}} + \chi^{\rm in}\label{eq:rho_i_thermal_pops} \\
    & \rho^{\rm in}_{\rm B} = \rho^{\rm fin}_{\rm th}(\beta) + \chi^{\rm fin} \equiv \frac{e^{-\beta H_{t_f}}}{Z_{f,\beta}}+\chi^{\rm fin}\,,\label{eq:rho_iB_thermal_pops}
\end{align}
where ${\rm tr}(\chi^{\rm in}) = {\rm tr}(\chi^{\rm fin}) = 0$, and the Hamiltonian $H$ of the system is not necessarily assumed as a time-independent operator. By substituting Eqs.\,(\ref{eq:rho_i_thermal_pops}) and (\ref{eq:rho_iB_thermal_pops}) into (\ref{detailed_be_app}) with $k=n=i$ and $\ell=m=f$, one finds 
\begin{equation}\label{eq:mimick_Jarzynski_1}
\frac{{\rm P}_{\Gamma}({\rm i,f})}{{\rm P}_{\widetilde{\Gamma}}({\rm f,i})} = \exp\left[\beta(\Delta E_{\rm i,f}-\Delta F)+\sigma^{\rm fin}_{\rm f}\left(\rho^{\rm in}_{\rm th}(\beta)\right) + \Sigma^{\rm fin}_{\rm f}\left(\chi^{\rm in}\right)-\widetilde{\sigma}^{\rm fin}_{\rm i}\left(\rho^{\rm fin}_{\rm th}(\beta)\right)-\widetilde{\Sigma}^{\rm fin}_{\rm i}\left(\chi^{\rm fin}\right)\right],
\end{equation}
where
\begin{eqnarray}
&&\sigma^{\rm fin}_{\rm f}\left(\rho^{\rm in}_{\rm th}(\beta)\right) \equiv \ln p^{\rm fin}_{\rm f}\left(\rho^{\rm in}_{\rm th}(\beta)\right), \\
&&\Sigma^{\rm fin}_{\rm f}\left(\chi^{\rm in}\right) \equiv \ln\left(1+\frac{p^{\rm fin}_{\rm f}(\chi^{\rm in})}{p^{\rm fin}_{\rm f}\left(\rho_{\rm th}^{\rm in}(\beta)\right)}\right), \\
&&\widetilde{\sigma}^{\rm fin}_{\rm i}\left(\rho^{\rm fin}_{\rm th}(\beta)\right) \equiv \ln\widetilde{p}^{\rm fin}_{\rm i}\left(\rho_{\rm th}^{\rm fin}(\beta)\right), \\
&& \widetilde{\Sigma}^{\rm fin}_{\rm i}\left(\chi^{\rm fin}\right) \equiv \ln\left(1+\frac{\widetilde{p}^{\rm fin}_{\rm i}(\chi^{\rm fin})}{\widetilde{p}^{\rm fin}_{\rm i}\left(\rho_{\rm th}^{\rm fin}(\beta)\right)}\right),
\end{eqnarray}
$p^{\rm fin}_{\rm f}(\mathcal{A}) \equiv \rm{tr}\left(\Pi^{\rm fin}_{\rm f}\Phi(\mathcal{A})\right)$, and $\widetilde{p}^{\rm fin}_{\rm i}(\mathcal{A}) \equiv \rm{tr}\left(\Pi^{\rm fin}_{\rm i}\widetilde{\Phi}(\mathcal{A})\right)$ with $\mathcal{A}$ a generic linear operator. Note that $p^{\rm fin}_{\rm f}(\mathcal{A})$ and $\widetilde{p}^{\rm fin}_{\rm i}(\mathcal{A})$ denote the probability to measure the ${\rm f}$-th and ${\rm i}$-th final energy values of the quantum system in the forward and backward process, respectively, conditioned to have evolved the thermal contribution of the initial state (without coherence terms in the energy eigenbasis). Finally, Eq.~(\ref{eq:mimick_Jarzynski_1}) can be simplified as
\begin{equation}\label{eq:mimick_Jarzynski_3_over_2_app}
    \frac{{\rm P}_{\Gamma}({\rm i,f})}{{\rm P}_{\widetilde{\Gamma}}({\rm f,i})} = \exp\left[\beta(\Delta E_{\rm i,f}-\Delta F)+\Delta\sigma_{\rm i,f}+\Delta\Sigma_{\rm i,f}\right]
\end{equation}
by introducing the quantities
\begin{equation}
    \Delta\sigma_{\rm i,f} \equiv \sigma^{\rm fin}_{\rm f}\left(\rho^{\rm in}_{\rm th}(\beta)\right) - \widetilde{\sigma}^{\rm fin}_{\rm i}\left(\rho^{\rm fin}_{\rm th}(\beta)\right)\,\,\,\,\,\,\,\,\text{and}\,\,\,\,\,\,\,\,\Delta\Sigma_{\rm i,f} \equiv \Sigma^{\rm fin}_{\rm f}\left(\chi^{\rm in}\right) - \widetilde{\Sigma}^{\rm fin}_{\rm i}\left(\chi^{\rm fin}\right).
\end{equation}
It is worth observing that, if no quantum coherence is present neither in $\rho_{0}$ nor in $\rho^{\rm in}_{\rm B}$ (i.e., $\chi^{\rm in}=\chi^{\rm fin}=0$), then $\Delta\Sigma_{\rm i,f} = 0$. Thus, $\Delta\Sigma_{\rm i,f}$ can be considered as a \emph{correction}, due to initial coherence in the energy basis of the system, to the entropy difference $\Delta\sigma_{\rm i,f}$ obtained by propagating initial thermal states in the forward and backward processes, respectively. 

The form of the detailed fluctuation theorem used in the main text is a particular case of Eq.\,(\ref{eq:mimick_Jarzynski_3_over_2_app}) where $\chi^{\rm fin}$ is assumed to vanish. This choice is motivated by our aim to consider the minimal modification to the ``Jarzynski set-up'', respect to which only coherence in the initial state of the forward dynamics is added.

\section{Part B: Derivation of the integral fluctuation theorem for $\Delta\Sigma$
}\label{app:IFT_DeltaSigma}
\renewcommand{\theequation}{B-\arabic{equation}}
\setcounter{equation}{0}

From Eq.~\eqref{eq:mimick_Jarzynski_3_over_2_app} the integral fluctuation theorem of Eq.~\eqref{eq_integral_fluct_relation} can be easily obtained. 

Instead, for what concerns the integral fluctuation theorem involving the sole coherence induced entropy production, let us consider the case in which $\chi^{\rm fin} = 0$. Upon substitution, one has
\begin{equation}
    \exp\left[-\Delta\Sigma_{\rm i,f}\right] = \frac{p^{\rm fin}_{\rm f}\left(\rho_{\rm th}^{\rm in}(\beta)\right)}{p^{\rm fin}_{\rm f}\left(\rho_{\rm th}^{\rm in}(\beta)\right)+p^{\rm fin}_{\rm f}(\chi^{\rm in})}\,.
\end{equation}
Thus, taking the average over the EPM probability distribution of the forward process $\Gamma$, we can conclude that 
\begin{equation}
    \left\langle\exp\left[-\Delta\Sigma_{\rm i,f}\right]\right\rangle_{\Gamma} = \sum_{\rm f}p^{\rm fin}_{\rm f}\left(\rho_{\rm th}^{\rm in}(\beta)\right) = 1 \,,
\end{equation}
where we have used the fact that $p^{\rm fin}_{\rm f}\left(\rho_{\rm th}^{\rm in}(\beta)\right)$ is itself a probability distribution normalized to $1$.

\section{Part C: Photodynamics of the NV-center}\label{app:OpenDyn}
\renewcommand{\theequation}{C-\arabic{equation}}
\setcounter{equation}{0}

\subsection{Theoretical modelling}

In our experimental set-up, the photodynamics of the NV center is well described by a seven-level quantum model~\cite{PhysRevA.88.020101,PhysRevB.74.104303,HernandezPRR2019}, which, however, can be effectively reduced to a two-level quantum system as detailed in the Supplemental Material of Refs.~\cite{HernandezPRR2019}. Here, for completeness of exposition, we report the modelling of this effective two-level open dynamics of the NV center, by resorting to the super-operator formalism~\cite{HavelJMP2003} and working in the energy eigenbasis. 

Let us thus model the NV center as a two-level quantum system subject to a dissipative dynamics. The dissipative dynamics is induced by the laser pulses and it is described by the linear super-operators ${\bf S}\in \mathbb{C}^{4\times 4}$ acting directly on the column vector ${\rm col}[\rho_t] \in \mathbb{C}^{4\times 1}$, with ${\rm col}[\rho]$ denoting the vectorization of the density operator $\rho \in \mathbb{C}^{2\times 2}$.  
The super-operator ${\bf S}$ is explicitly given by
\begin{equation}
{\bf S} = \frac12
\begin{pmatrix}
2 - p_{\rm abs}(k_c - p_{d}\cos\alpha) & \displaystyle{p_{\rm abs}k_{sc}} & \displaystyle{p_{\rm abs}k_{sc}} & p_{\rm abs}(p_d\cos\alpha+k_c) \\
p_{\rm abs}(k_{sc} + p_d\sin\alpha) & 2-\displaystyle{p_{\rm abs}(1+k_s)} & -\displaystyle{p_{\rm abs}(k_s-1)} & p_{\rm abs}(p_d\sin\alpha-k_{sc}) \\
p_{\rm abs}(k_{sc} + p_d\sin\alpha) & -\displaystyle{p_{\rm abs}(k_s - 1)} & 2-\displaystyle{p_{\rm abs}(1+k_s)} & p_{\rm abs}(p_d\sin\alpha-k_{sc}) \\
p_{\rm abs}(k_c - p_d\cos\alpha) & -\displaystyle{p_{\rm abs}k_{sc}} & -\displaystyle{p_{\rm abs}k_{sc}} & 2-p_{\rm abs}(p_d\cos\alpha+k_c)
    \end{pmatrix},
\end{equation}
where
\begin{equation}
k_c = 1 - (1 - p_d)(\cos\alpha)^2,\qquad k_s = 1 - (1 - p_d)(\sin\alpha)^2,\qquad k_{sc} = (1-p_d)\sin\alpha\cos\alpha,
\end{equation}
$p_{\rm abs}$ is the absorption probability, $p_d$ is the probability of population transfer to $|0\rangle_S$, and $\alpha\in [0,\pi/2]$. 

The unitary dynamics in between two consecutive pulses is instead described by the linear operator 
\begin{equation}
    {\bf U} = \exp\left(-i\tau\left(H\otimes\mathbbm{I}_{2} - \mathbbm{I}_{2}\otimes H^{\ast}\right)\right)
\end{equation}
with $\hbar$ set to $1$, $\mathbbm{I}_{2}$ denoting the $2\times 2$ identity matrix, and $H = \omega\,\sigma_{z}/2$ the Hamiltonian of the effective two-level system. 

Considering a total number of pulses $N$ we thus have
\begin{equation}
    {\rm col}[\rho_t] = {\bf L}\,{\rm col}[\rho_0] \quad \text{with} \quad {\bf L} \equiv \left({\bf S}{\bf U}\right)^{N}.
\end{equation}
For a fixed set of values of the parameters $(p_{\rm abs},p_d,\alpha,\tau,\omega)$, the super-operator ${\bf L}$ governing the open dynamics of the NV center possesses a unique steady-state $\rho^*$ which is characterised by a non vanishing coherence in the energy basis.

The Kraus representation for ${\bf L}$ can be uniquely determined by diagonalizing its Choi matrix 
\begin{equation}
    {\bf T} \equiv \sum_{k,j=0}^{1}\left(E_{kj}\otimes\mathbbm{I}_{2}\right){\bf L}\left(\mathbbm{I}_{2}\otimes E_{kj}\right) = \sum_{\ell=0}^{3}\xi_{\ell}{\bf u}_{\ell}{\bf u}_{\ell}^{\dagger}
\end{equation}
where $E_{kj} \equiv |k\rangle\!\langle j|$, with $|0\rangle \equiv (1,0)^{T}$ and $|1\rangle \equiv (0,1)^{T}$, and $(\xi_{\ell},{\bf u}_{\ell})$ denotes the $j$-th pair (eigenvalue, eigenvector) resulting from the eigenvector decomposition of ${\bf T}$. The Kraus operators $\{K_{\ell}\}$ associated with the open map ${\bf L}$ are thus implicitly provided by the following relation:
\begin{equation}\label{eq:Kraus_operators}
    {\rm col}\left[K_{\ell}\right] = \sqrt{\xi_{\ell}}{\bf u}_{\ell} \quad \text{such that} \quad \rho_{t} = \sum_{\ell=0}^{3}K_{\ell}\,\rho_{\rm i}K_{\ell}^{\dagger}\,.
\end{equation}
One can easily determine that the open dynamics of the NV center is correctly described by three Kraus operators $K_{\ell}$. These Kraus operators can then be considered for the derivation of the backward dynamics as discussed in the main text. 

\subsection{Experimental data and simulation}

\begin{figure}
\centering
\includegraphics[width=0.5\columnwidth]{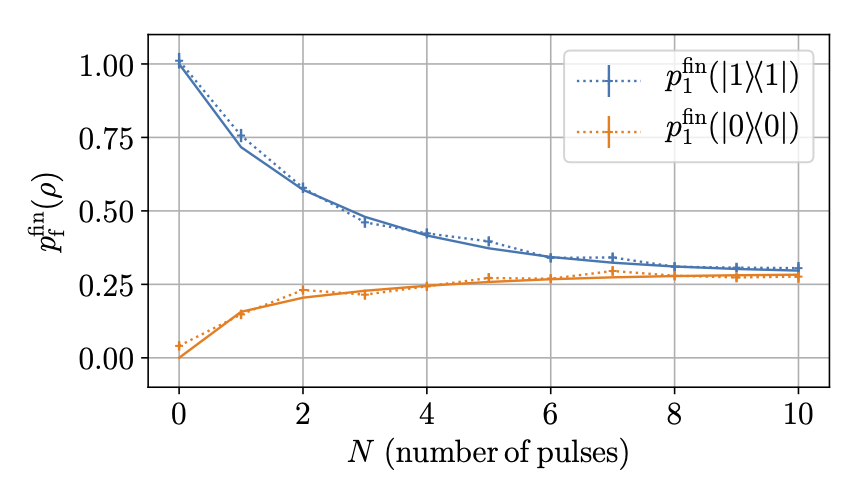}%
\includegraphics[width=0.5\columnwidth]{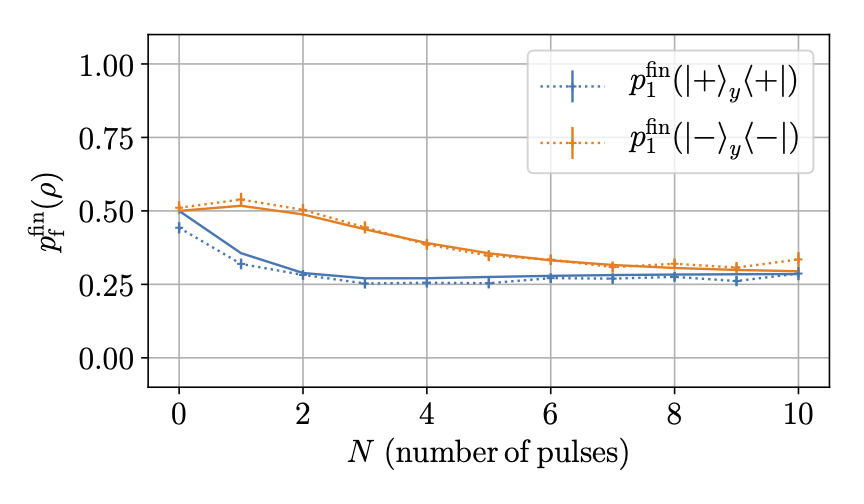}
\caption{
Measurements of the EPM  protocol as a function of the number $N$ of pulses for the initial pure states $\{|0\rangle,|1\rangle,|+\rangle_y,|-\rangle_y\}$. The crosses with error bars are the experimental data (dotted line is a guide to the eye). Solid lines represent the numerical simulation of the dynamics described in the previous section, using the parameters $p_{\rm abs}=0.700$, $p_d = 0.255$. The other parameters are, as mentioned in the main text, $\alpha = \pi/4$ (i.e., $\delta=-\Omega$), $\tau\omega\simeq (2\pi)0.9$, and $\tau=190$~ns.
}
\label{figSupp:direct_measurements}
\end{figure}

As described in the main text, in our experiments we measure the EPM probability $p^{\rm fin}_{\rm f}(\rho)$, i.e., the probability of obtaining $E^{\rm fin}_{\rm f}$ when measuring the energy of the system at the final time $t_{\rm f}=N\tau$ (where $N$ is the number of laser pulses and $\tau$ is the time between them), assuming that the system is initialized into the state $\rho$. In particular, we performed four independent experiments for each of the four initial states: $|0\rangle$, $|1\rangle$, $|+\rangle_y$, and $|-\rangle_y$. The results of such measurements are shown in Fig.~\ref{figSupp:direct_measurements}. 
Notice that we only measure the probability $p^{\rm fin}_{{\rm f}=1}$ associated to the final excited state $|1\rangle$. This is because, for a two level system, the remaining probability is obtained as $p^{\rm fin}_{{\rm f}=0} = 1 - p^{\rm fin}_{{\rm f}=1}$. 
As mentioned in the main text, the classical mixtures we are interested in are a convex combination of these four pure states. For example, the EPM probability associated with the initial thermal state $\rho^{\rm in}_{\rm th} = e^{-\beta H_{t_{i}}}/Z_{i}$ is obtained as $p^{\rm fin}_{\rm f}(\rho^{\rm in}_{\rm th}) = p(E_0^{\rm in}) p^{\rm fin}_{\rm f}(|0\rangle\langle0|) + p(E_1^{\rm in}) p^{\rm fin}_{\rm f}(|1\rangle\langle1|)$, where $p(E_{\rm i}^{\rm in}) = e^{-\beta E_{\rm i}}/Z_{i}$.
Similarly, the probability $p^{\rm fin}_{\rm f}(\rho_0)$ for a given initial state (see main text) $\rho_0 = p |1\rangle\!\langle 1| + (1-p)|+\rangle_y\!\langle +|$, with $p\in [0,1]$, is obtained as $p^{\rm fin}_{\rm f}(\rho_0) = p p^{\rm fin}_{\rm f}(|1\rangle\!\langle 1|) + (1-p)p^{\rm fin}_{\rm f}(|+\rangle_y\!\langle +|)$. 

In Fig.~\ref{figSupp:direct_measurements} we compare the experimental data with the numerical simulation of the dynamics, using the model described in the previous section. The values of the parameters $p_{\rm abs}$, and $p_d$ are selected by minimizing the sum of the squares of the residuals between data and simulation. 

\begin{figure}
\centering
\includegraphics[width=\columnwidth]{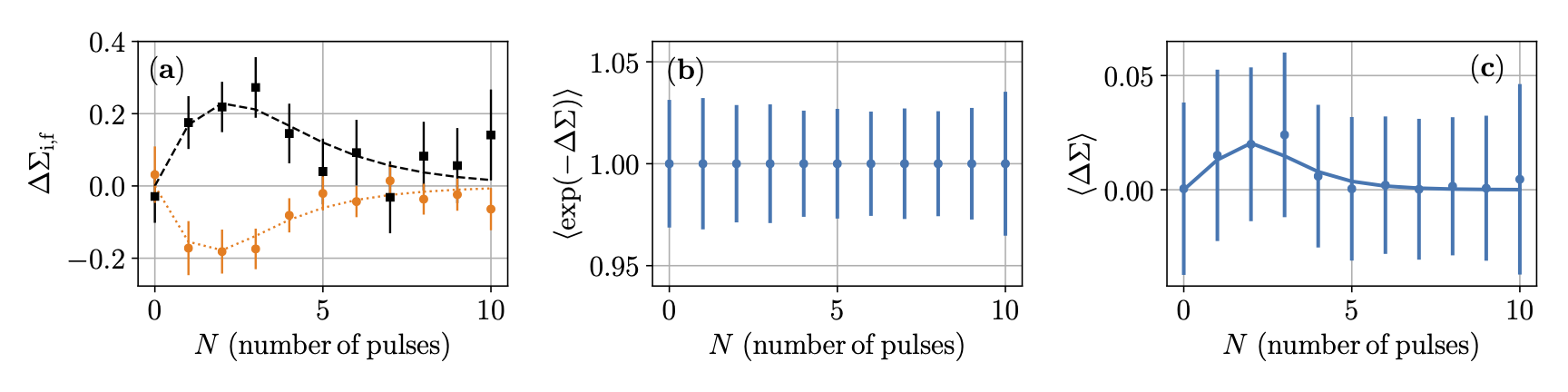}
\caption{Panel {\bf (a)}: Experimental values (markers with error bars) of the irreversible entropy production. Black squares are $\Delta\Sigma_{1,1}=\Delta\Sigma_{0,1}$, and orange bullets are $\Delta\Sigma_{1,0}=\Delta\Sigma_{0,0}$. Dashed black line and dotted orange line are the corresponding numerical simulations. Panel {\bf (b)}: Experimental verification of the fluctuation theorem $\left\langle e^{-\Delta\Sigma}\right\rangle_{\Gamma}=1$, Eq.~\eqref{SigmaFT}, for the coherence-affected irreversible entropy production, as a function of the number $N$ of pulses used to drive the dynamics of the NV center. Panel {\bf (c)}: Average experimental coherence-affected entropy production as a function of $N$ (blue circles). In all panels, the experimental data are plotted against the predictions from the numerical simulations that are obtained by taking $|-\rangle_{y}=(\ket{0}-i\ket{1})/\sqrt{2}$ as the initial quantum state.}
\label{figSupp:DeltaSigma_and_meanEntropy}
\end{figure}

In Fig.~2 of the main text we show the results of the irreversible entropy production $\Delta\Sigma$, obtained for an initial state $|+\rangle_y$. In Fig.~\ref{figSupp:DeltaSigma_and_meanEntropy} we present similar results but for an initial state $|-\rangle_{y}=(\ket{0}-i\ket{1})/\sqrt{2}$. 

\end{document}